# Energy Transfer within the Hydrogen Bonding Network of Water Following Resonant Terahertz Excitation

Hossam Elgabarty[1], Tobias Kampfrath[2,3], Douwe Jan Bonthuis[3], Vasileios Balos[2], Naveen K. Kaliannan[1], Philip Loche[3], Roland R. Netz[3], Martin Wolf[2], Thomas D. Kühne[1]* and Mohsen Sajadi[2]*

[1]*Department of Chemistry, University of Paderborn, Paderborn, Germany*

[2]*Fritz-Haber-Institut der Max-Planck-Gesellschaft, Berlin, Germany*

[3]*Department of Physics, Freie Universität Berlin, Berlin, Germany*

\* tdkuehne@mail.uni-paderborn.de, sajadi@fhi-berlin.mpg.de.

**Energy dissipation in water is very fast and more efficient than in many other liquids. This behavior is commonly attributed to the intermolecular interactions associated with hydrogen bonds. Here, we investigate the dynamic energy flow in the hydrogen-bond network of liquid water by a pump-probe experiment. We resonantly excite intermolecular degrees of freedom with ultrashort single-cycle terahertz pulses and monitor its Raman response. By using ultrathin sample-cell windows, a background-free bipolar signal whose tail relaxes mono-exponentially is obtained. The relaxation is attributed to the molecular translational motions, using complementary experiments and force-field and *ab initio* molecular dynamics simulations. They reveal an initial coupling of the terahertz electric field to the molecular rotational degrees of freedom whose energy is rapidly transferred, within the excitation pulse duration, to the restricted-translational motion of neighboring molecules. This rapid energy transfer may be rationalized by the strong anharmonicity of the intermolecular interactions.**

Water is a major substance on the earth surface. Its diverse anomalous properties make life on our planet viable. Notably, its large heat capacity turns oceans and seas into giant heat reservoirs for regulating the earth climate. In living organisms, the same property makes water a superb thermal buffer for the function of bio-chemical reactions[1,2,3]. These thermodynamic peculiarities are commonly attributed to water's ability to form an intermolecular complex network which is based on thermally fluctuating hydrogen (H) bonds. Interestingly, as each water molecule forms on average close to four H-bonds with ~1ps lifetime in an almost tetrahedral configuration,[4,5,6] the three-dimensional network of H-bonded water molecules encompasses complex collective/cooperative intermolecular degrees of freedom with a very diverse dynamics[7].

The molecular dynamics associated with this network, including the restricted translations and rotations and also the diffusive motions, cover an exceptionally broad frequency range, with a bandwidth of more than 1000 cm$^{-1}$. These spectrally broad intermolecular degrees of freedom may then serve as a heat sink with abundant pathways for the accommodation/dissipation of deposited excess energy in water,[8] explaining its large heat capacity[9]. The extension of the intermolecular modes to high frequencies makes it also an ideal/efficient thermal bath for ultrafast relaxation of energy from vibronically hot (bio-) molecules, thereby avoiding their permanent thermal damage[10,11,12]. To elucidate the molecular mechanism of the energy dissipation in water and understand the role of collective intermolecular motions in this process, the time scale of energy dissipation and the strength of intermolecular interactions should be determined by experiments.

While linear-type spectroscopic methods, such as dielectric relaxation, determine the polarization decay of the infrared-active modes of liquids, nonlinear infrared spectroscopy has extensively been used to provide complementary microscopic insights into the accompanying energy dissipation processes. For example, the O-H stretch vibration has been used as a local probe to interrogate the dynamics of its surrounding[14,15,16,17]. Previous studies using this approach proposed the dipole-dipole interaction to be the main mechanism of the vibrational energy transfer in water[15] and determined the time scale of this process to be <100fs[16]. Moreover, ultrafast (sub-100 fs) energy transfer from the OH-bending vibration to the librational (hindered rotational) motion has also been resolved[18,19,20].

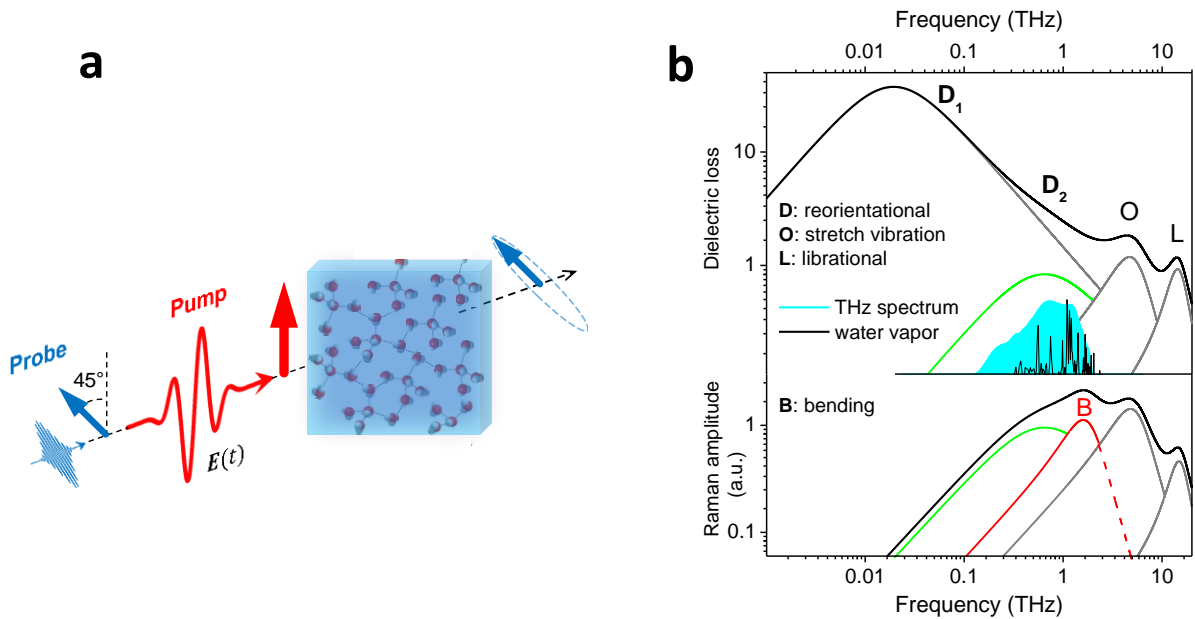

**Figure 1 | Dynamic THz Kerr effect. a**, An intense THz pump pulse is used to induce optical birefringence in water, which is monitored by an optical probe pulse that becomes elliptically polarized upon traversing through the medium. **b**, Equilibrium dielectric loss (Im $\varepsilon$) and incoherent Raman spectra of water.[13] The black sharp lines are the THz absorption lines of water vapor and the spectrum of the excitation THz field is indicated by the cyan area.

However despite these efforts, there are still various open questions regarding the energy flow in the H-bond network. For example, to what extent do intermolecular modes and processes contribute to the energy transfer within the H-bond network of water? What is the time scale for the energy transfer between these motions, and how strongly they are coupled? We believe that a more accurate understanding of the energy dissipation process in water will emerge by direct interrogation of the intermolecular degrees of freedom. Since, the spectral fingerprint of the intermolecular H-bonding dynamics lies in the terahertz (THz) frequency range, it is promising to resonantly pump the low-frequency collective modes/processes of water with a THz pulse and to probe the response of the system in real time.

This method has already provided insights into intramolecular mode coupling in halogenated liquids,[21] into the resonant coupling of THz radiation to permanent molecular dipoles in various polar liquids[22] and has indicated that the response of water cannot be described by Langevin-type molecular rotational dynamics[23]. In principle, this method enables us to assign the original energy recipient mode, the time scale and the pathways of energy flow into intermolecular degrees of freedom. Such experiments may eventually map out the complex energy potential surface of the H-bonded network of water and ultimately enable us to model the structural dynamics of water[24].

Here, we resonantly excite the collective rotational degrees of freedom of water with an intense THz pulse and probe the resulting optical anisotropy in a THz Kerr effect (TKE) configuration[25,26]. To reveal the origin of the resulting response, we perform complementary experiments, including the temperature-dependent TKE of liquid water, TKE of water vapor and the optical Kerr effect (OKE) of liquid water. We also perform both the *ab initio* molecular dynamics (AIMD) simulations and the force field-based MD (FFMD) simulations under the effect of the same THz field to gain deeper insights into the process of intermolecular energy transfer in water. In the AIMD simulations the interatomic interactions are computed "on-the-fly" by electronic structure calculations.

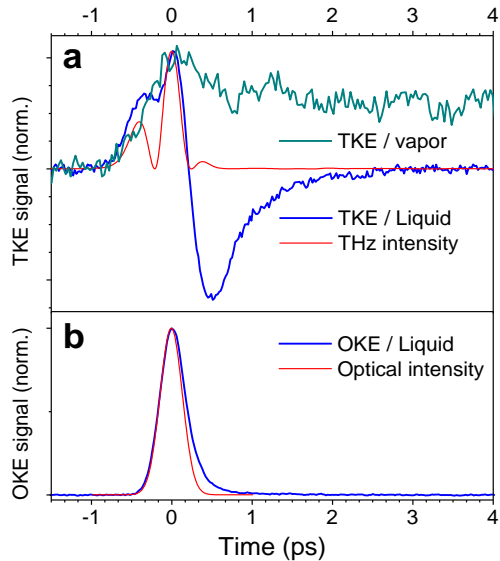

**Figure 2 | Dynamic Kerr effect of water. a**, THz-field-induced transient optical birefringence of water vapor (green line) and liquid water (blue line) and the square of the driving THz field (red line). **b**, Optical Kerr effect of water at (blue line). Red line shows the intensity of optical pump pulse estimated, as a Gaussian fit to the left flank of OKE response of water. Liquid measurements are performed at 22°C.

## Results

**Experimental setup**. A schematic of the TKE experiment is shown in **Fig. 1a**. An intense linearly polarized THz electric field (peak strength of ~2 MV/cm) excites water (double-distilled[27]). Its amplitude spectrum is shown in **Fig. 1b** (blue area), whereas its instantaneous intensity $E_{\text{THz}}^2(t)$ is shown in **Fig. 2a** (red solid line). The pump- induced optical birefringence $\Delta n(t)$ is measured by a probe pulse (800 nm, 2 nJ, 8 fs) whose linear polarization acquires ellipticity by traversing the sample. Their dielectric loss (Im $\varepsilon$) and the incoherent Raman spectrum of liquid water are shown in **Fig. 1b**. The double-distilled liquid water film (thickness of 100 µm) is held between a rear glass window and a 150 nm thick silicon nitride (SiN) membrane as the entrance window[28]. These thin windows exhibit a negligibly small Kerr signal[16]. In contrast, ~1 mm thick windows used in a previous study cause a large background signal, whose separation from the liquid response is technically challenging[23]. As seen in **Fig. 1b**, the THz pump spectrum overlaps with the rotational transitions of water vapor (black spikes).

The THz-field-induced optical birefringence of liquid water is compared to that induced by an optical pump pulse. Both optical and THz excitation are conducted in the same setup under otherwise identical conditions. The intensity envelopes of the optical and the THz pump pulses have approximately the same temporal width of about 350 fs, thereby allowing for straightforward comparison of the TKE and the OKE data.

**TKE of liquid water**. The pump-probe signals for various samples and excitation conditions are shown in **Fig. 2**. The TKE signal of water vapor (green line) is unipolar and decays over hundreds of picoseconds. Strikingly and in contrast, the TKE signal of liquid water (blue line) is bipolar and relaxes within a few picoseconds. The signal does not change sign when the THz field is reversed, in line with the quadratic dependence of the TKE signal on the THz field amplitude (Fig. S1). The transient birefringence of water following optical excitation is shown in **Fig. 2b**. Analogous to the TKE signal of water vapor and in line with previous studies[29,30,31,32,33] the OKE signal is unipolar: a signal with a spike around time zero (the instantaneous electronic response) and a relatively weak relaxation tail. A comparison between these results reveals three distinct features of the TKE signal of liquid water.

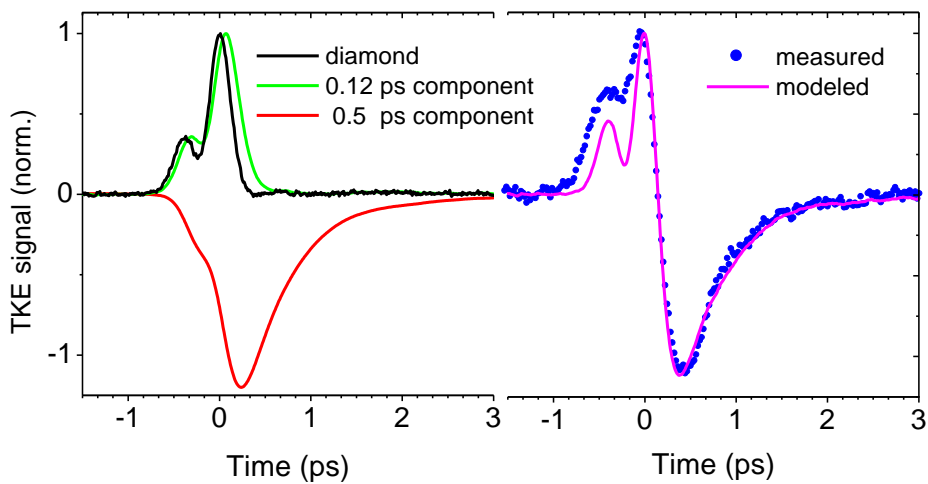

**Figure 3 | Dynamic components of the TKE signal of water.** left, Square of THz field (black line) is convoluted with two exponential functions with $\tau_1 \approx 0.12$ ps (green line) and $\tau_2 \approx 0.5$ ps (red line). right, The Sum of the latter two curves gives rise to the magenta line, which captures almost all features of the measured TKE signal of water (blue dots).

(i) *Bipolarity*. In stark contrast to the TKE signal of water vapor and the OKE signal, the TKE response of liquid water is bipolar. The bipolar TKE signals of liquids have so far been observed only in water and n-alcohols[34,22]. Zalden et al. using pulses with ~0.2 THz center frequency, observed bipolar TKE signals for n-alcohols[23]. Our TKE experiment of alcohols, however revealed unipolar signals after excitation with pulses with ~1 THz center frequency, suggesting that the polarity of the TKE signal depends on the nature of the excited mode/process in H-bonded liquids (see Fig. S2).

(ii) *Relaxation*. The tail of the TKE signal relaxes with a time constant of ~0.5ps. To determine this time constant, we phenomenologically modeled the TKE signal of water by convoluting two exponential functions with the assumed instantaneous electronic response of water. The latter is estimated by the TKE signal of a thin diamond plate (black line in **Fig. 3**). As illustrated in **Fig. 3**, two exponential components with opposite signs and decay times of ~0.12ps (green line) and ~0.5ps (red line) can reproduce the experimental result reasonably well (magenta line). The discrepancy at the leading edge of the THz pulse most likely arises from the dispersion of water which was neglected in the modeling (see Methods Section). The 0.5ps component has also been reported in previous OKE studies of water[29,30,31,32,33]. As the faster 0.12ps component overlaps with the instantaneous electronic response, we focus in the following on the 0.5ps component. Note also that, as we use an ultrathin cell window (150 nm thick SiN membrane), the measured TKE signal can be uniquely assigned to the liquid. For thick windows, subtraction of the window response from the liquid response is essential and often a technical challenge which may easily lead to the extraction of different relaxation time constants from the measured signal[16].

(iii) *Enhancement*. Relative to their input energies, the nuclear portion of the TKE signal of water has an enhanced amplitude compared to that of the OKE signal. The input energies are calibrated via the amplitude of the instantaneous electronic response of water for both the THz and the optical excitations. To estimate the magnitude of the enhancement, both signals are normalized to the peak of their electronic responses. As both pump pulses have a comparable duration, this procedure is tantamount to the normalization of the signals to their corresponding driving pulse energies[22]. Accordingly, by comparing the amplitude of the signals at the time delay of 1ps (when the electronic responses are fully over) we obtain a TKE versus OKE signal enhancement factor of ~30.

## Discussion

**Optical birefringence.** Due to the action of the pump field (polarized along $x$, see **Fig. 1a**), the probe pulse (polarized at 45° relative to $x$) encounters a transient difference $\Delta n = n_x - n_y$ between the refractive indices along the $x$ and $y$ directions. The resulting birefringence is given by[35]

$$\Delta n = \Delta n_{\text{rot}} + \Delta n_{\text{trans}} \propto \langle \Delta \Pi_{xx}^{(m)} - \Delta \Pi_{yy}^{(m)} \rangle, \qquad (1)$$

where $\langle . \rangle$ denotes averaging over all molecules $m$, and $\Delta \Pi_{ij}^{(m)}$ is the pump-induced change in the polarizability tensor $\mathbf{\Pi}^{(m)} = \Pi_{ij}^{(m)}$ of molecule number $m$. Note that $\mathbf{\Pi}^{(m)}$ refers to the liquid phase and contains contributions from interactions/collisions with other molecules, in contrast to a single gas-phase molecule. The variation $\Delta \mathbf{\Pi}^{(m)}$ can, in principle, be written as a sum $\Delta \mathbf{\Pi}_{\text{rot}}^{(m)} + \Delta \mathbf{\Pi}_{\text{trans}}^{(m)}$ whose two contributions arise, respectively, from pump-induced rotation of the molecules and/or changes in all other degrees of freedom, that is, intra- or intermolecular translational coordinates.

Averaging $\Delta \mathbf{\Pi}_{\text{rot}}^{(m)}$ over all molecules according to Eq. (1) yields an expression for $\Delta n_{\text{rot}}$ that scales with the degree of molecular alignment $\langle P_2(\cos\theta) \rangle = \langle 3 \cos^2 \theta^{(m)} - 1 \rangle / 2$ and the molecular polarizability anisotropy $\Delta \Pi_{\text{rot}}^{(m)}$ i.e. $\Delta n_{\text{rot}} \propto \Delta \Pi_{\text{rot}}^{(m)} \langle P_2(\cos\theta) \rangle$. The latter characterizes the degree of anisotropy of the unperturbed $\mathbf{\Pi}^{(m)}$ and is usually labeled $\Delta\alpha$ in the case of single molecules[44,35].

The averaged $\Delta \mathbf{\Pi}_{\text{trans}}^{(m)}$ makes another contribution $\Delta n_{\text{trans}}$ to the transient birefringence $\Delta n$ and arises from directly or indirectly pump-induced changes in the translational rather than rotational degrees of freedom. In the following, we discuss which of the two contributions can explain the TKE signal of water.

**THz induced molecular orientation and alignment**. To determine the origin of the TKE response, one needs to resolve the mechanism of the THz field interaction with water. To this end, we use molecular dynamics (MD) simulations and calculate the degree of molecular orientation as an ensemble average of the angle between the THz electric field and the molecular bisector. The results of AIMD and FFMD simulations are given in **Fig. 4a.** They both show a discernible orientation $\langle \cos\theta \rangle$ of water molecules whose patterns follow the THz wave form with a small phase shift.

To ensure that THz electric field primarily drives the rotational degree of freedom, we also obtain $\langle \cos\theta \rangle$ from the experimentally determined dielectric susceptibility of water. The latter in the region that overlaps with THz pulse spectrum can be fit with two Debye processes[36,37] and commonly is attributed to the rotational degrees of freedom[37,38,39,40,41]. Interestingly, as detailed in Fig. S3 and supplementary Note 1, the THz electric field induced polarization $P(\omega) = \varepsilon_0 \chi(\omega) E(\omega)$ leads to a comparable degree of molecular orientation to that obtained from MD simulations, implying that the THz pulse spectrum overlaps with the molecular rotational degrees of freedom.

This finding is also endorsed by the enhanced TKE vs. OKE signal of water. Our previous TKE studies of polar liquids corroborate that THz electric coupling to the rotational degrees of freedom leads to an enhanced TKE vs. OKE signal, while no enhancement is observed after resonant excitation of the (restricted) translational motions[22,42].

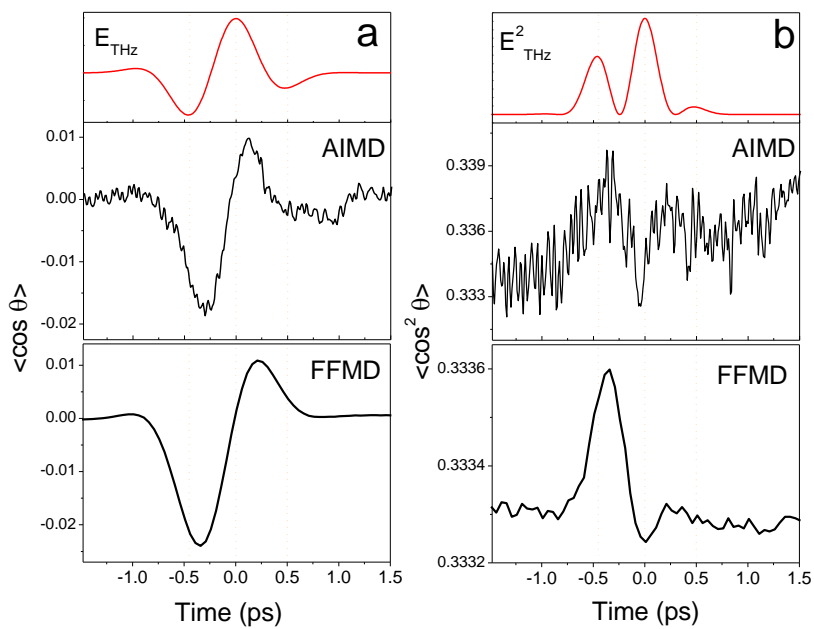

**Figure 4 | Molecular orientation and alignment.** The orientational dynamics of water molecules after THz excitation obtained from both the AIMD and FFMD simulations. The θ in y-axis is the angle between the water bisector and the THz electric field polarization axis for **a**, the molecular orientation and **b**, the molecular alignment. The angle brackets indicate the ensemble averages. Top panels show the THz electric field and the THz intensity. Note that, FFMD simulations give rise to the results with much higher signal-to-noise ratio, as 5000 trajectories are averaged for 5360 water molecule per trajectory.

As $\Delta n_{\text{rot}}$ scales with molecular alignment $\langle \cos^2 \theta - 1/3 \rangle$, we also simulated the degree of molecular alignment. This simulation can clarify whether the TKE response of water originates from the reorientational relaxation of single molecules. Notably, as shown in **Fig. 4b**, the temporal pattern of $\langle \cos^2 \theta \rangle$ from both AIMD and FFMD simulations manifests an ultrafast dynamics of a rise and a decay almost fully within the temporal duration of the THz intensity profile and lacks the relaxation tail of the TKE signal. Moreover, $\langle \cos^2 \theta \rangle$ gives rise to a $\Delta n_{\text{rot}}$ ($\propto \Delta\alpha \langle P_2(\cos\theta) \rangle$) which is about two orders of magnitude larger than the resolved $\Delta n$ in the TKE experiment (see Supplementary Note 1). These discrepancies suggest that the reorientational relaxation of single water molecules does not make a dominant contribution to the TKE signal of water.

**THz Kerr effect and dielectric response of water**. In addition to the argument given above, we also consider whether the expected Raman response of the Debye processes of water can explain the TKE response of water. The dynamic Kerr response of the slowest Debye process of water (**Fig. 1b**) with $\tau_{D_1} \approx$ 9ps[13] is expected to relax with a time constant of $\tau_{D_1}/3 \approx$ 3ps, or longer[43]. However, the TKE relaxation is about 6 times faster than the expected dynamic Kerr relaxation of the $D_1$ process. Thereby, it is unlikely that the TKE response of water originates directly from this process.

The fast Debye process $D_2$ is commonly attributed to the reorientational motion of under-coordinated single water-type molecules[36,37]. However, the reported dielectric relaxation time of this process is very diverse, ranging from $\tau_{D_2} \sim$ 1.2ps[37] to ~0.25ps[36], rendering the comparison between the dielectric and the Raman relaxation times less fruitful. In the following, using the properties of single water molecule, including the polarizability anisotropy, the dipole moment and also the bipolarity of the TKE response of liquid water, we discuss whether the $D_2$ process or single molecule reorientational dynamics can be the direct origin of the observed TKE signal of water.

(i) In the TKE process of polar liquids, the molecular dipole is aligned toward the electric field polarization axis. For this partially aligned molecular ensemble, we have recently provided arguments that $\Delta\alpha$ approximately equals the difference between the polarizability component along the molecular dipole axis and the polarizability average of the components in the plane perpendicular to the dipole axis[44].

(ii) The resulting optical birefringence in an ensemble of partially aligned molecules scales with $\Delta\alpha\langle P_2(\cos\theta)\rangle$[35]. If the molecular alignment is induced by the THz electric field ($E$) torque on the permanent dipoles, the alignment factor is given by $\langle P_2(\cos\theta)\rangle \propto \langle E \cdot \chi^{ori} * E\rangle$, where $\chi^{ori}(t)$ is the contribution of the orientational molecular motion to the total dielectric susceptibility of the liquid and asterisk $*$ denotes convolution[22]. As such, $\Delta n_{rot}$ in the TKE process scales linearly with $\Delta\alpha$. Therefore, the unipolar TKE signal of water vapor (see **Fig. 2a**) implies that $\Delta\alpha_{TKE}$ of a single water molecule is positive. Note that, in the OKE process $\langle P_2(\cos\theta)\rangle \propto \langle \Delta\alpha\, E^2\rangle$ and $\Delta n_{rot}$ scales quadratically with $\Delta\alpha$. Thus the OKE signal is always positive.

(iii) Mukamel et al. showed that the Raman spectrum of water can be reproduced with a polarizable model of the molecules. In their model, the polarizability of molecules is more anisotropic than that in the gas phase[45,46,47]. In a more recent *ab initio* theoretical study, Lu et al. considered the effect of the charge transfer and the local field of the hydrogen-bond network. They also showed that relative to the gas phase, the polarizability anisotropy of liquid water becomes more anisotropic[48]. In both studies the polarizability along the dipole axis is larger than the average of the two other components, such that $\Delta\alpha_{TKE} > 0$. Accordingly, the reorientational Kerr response of single water molecule $\Delta n_{rot}$ is expected to be resolved as a unipolar (positive) signal, in contrast to the conclusion drawn in Ref. 23. In the following, we provide evidence that the TKE response of water manifests the translational relaxation of water molecules.

**Energy dissipation**. Here, we refer again to our MD simulations and calculate the kinetic energy (KE) evolution of water molecules (See **Fig. 5**). In the AIMD simulations, we partition the total KE of the system into rotational $KE_{rot}$, translational $KE_{trans}$, and intramolecular vibrational components. Remarkably, the AIMD simulations show a rise of $KE_{trans}$ and a concomitant decay of $KE_{rot}$ with comparable amplitude. They reach to their peak values at a pump-probe delay of about 0.5 ps and relax to their equilibrium value exponentially with time constant $\tau \approx 0.75$ ps. Interestingly, this KE evolution directly declares the exchange of energy between the rotational and the translational degrees of freedom[49,50].

In the FFMD, the initial rise of the $KE_{trans}$ and its mono-exponential decay $\tau \approx 0.5$ ps is also observed (nicely comparable to the TKE relaxation tail and KE evolution in AIMD). However, $KE_{rot}$ shows only a monotonic rise until it reaches its new equilibrium value. As in FFMD the signal to noise ratio is high, the small temperature rise in the system can be observed as a step-like change in the KE of the system before and after the THz excitation.

The difference between the $KE_{rot}$ evolution in AIMD and FFMD may be related to the nonpolarizable/rigid model of water used in the FFMD simulations, a subject that is beyond the scopes of this work and will be scrutinized in a forthcoming publication.

The preceding MD simulation results and the features of the TKE signal of water strongly suggest a coupling between the IR-active rotational degrees of freedom and the low-frequency Raman-active translational motions. As shown schematically in **Fig. 6a**, two electric-field interactions with rotational transitions initially induce a rotational anisotropy (molecular alignment) in water. Due to rotational-translational (Ro-Trans) coupling (see Methods the "anharmonic coupling" section), the THz energy is transferred rapidly (faster than our time resolution ~350 fs) into the restricted-translational motions. Finally, the resulting anisotropy in the translational motion relaxes with ~0.5ps.

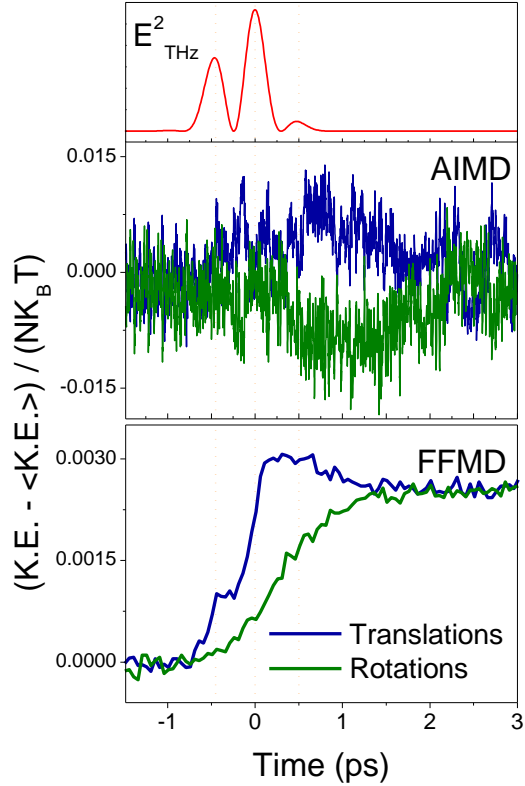

**Figure 5 | Molecular kinetic energy.** The relative Kinetic energy (KE) of the translational (blue) and rotational (green) motions after THz excitation obtained from both the AIMD and FFMD approaches. Top panel shows the THz intensity.

**Transient birefringence and rotational-translational coupling.** We now suggest a plausible microscopic scenario to connect the intermolecular rotational-translational coupling observed in the MD simulations to the THz Kerr effect of water. As discussed above, a prerequisite to observe a bipolar TKE signal is a negative molecular polarizability anisotropy. As this is not fulfilled for a single water molecule we seek supramolecular conformations whose defined polarizability anisotropy can be negative.

As highlighted in **Fig. 6b**, in the smallest conformation, i.e. a dimer one may realize an arrangement of a partial O-H dipole and the translational motion of its adjacent water molecule in an almost perpendicular configuration. The THz excitation causes a rotational torque on the O-H dipole. Because of the hydrogen binding between the two molecules, the latter torque bends the H-bonds[13] and transfers the deposited THz energy into the second water molecule and increases its translational kinetic energy. The latter motion which is restricted by the collision to other neighbors in the first solvation shell of the dimer, causes the deformation of the electronic cloud of the molecule and provides an additional polarizability component in the direction perpendicular to the O-H dipole. Accordingly, the polarizability anisotropy of the dimer becomes negative i.e. $\Delta \mathbf{\Pi}_{\text{trans}}^{(m)} = \mathbf{\Pi}_\parallel^{(m)} - \mathbf{\Pi}_\perp^{(m)} < 0$, as the polarizability of the dimer along the O-H dipole $\mathbf{\Pi}_\parallel^{(m)}$ is smaller than its perpendicular component $\mathbf{\Pi}_\perp^{(m)}$. As a result, the TKE response originating from the dimer i.e. $\Delta n_{\text{rot-dimer}} \propto \Delta \mathbf{\Pi}_{\text{trans}}^{(m)} \langle P_2(\cos \theta_{\text{dimer}}) \rangle$ becomes negative.

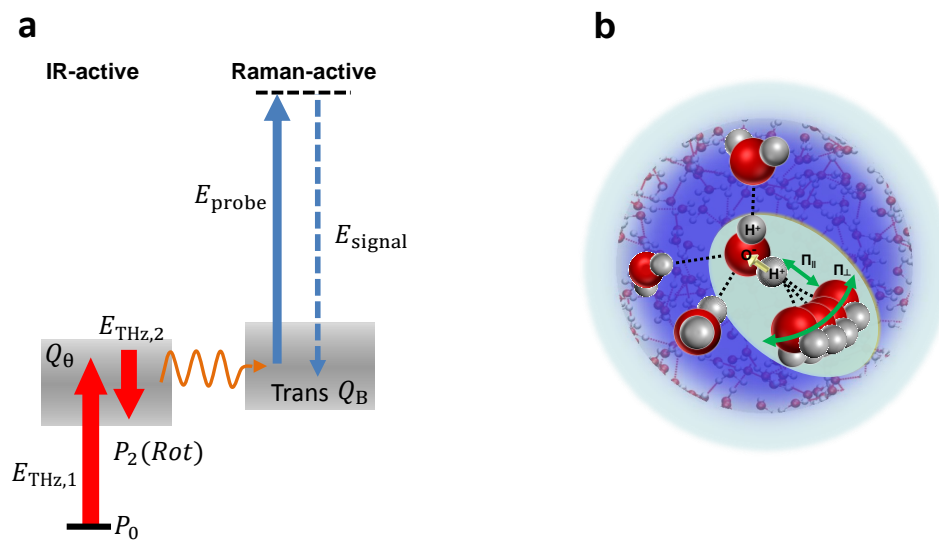

**Figure 6 | Intermolecular mode coupling in water. a**, Two-field interaction (red arrows) changes the rotational distribution of molecules (coordinate $Q_\theta$) from a disordered ($P_0$) into a (partially) aligned molecular system ($P_2$). The energy/momentum of this interaction is transferred into the restricted translational motion (coordinate $Q_B$) of the neighboring molecules, shown as a wavy arrow, because of the anharmonicity of the interaction energy between $Q_\theta$ and $Q_B$ (see Methods section for details). **b**, In water a supramolecular structure with a permanent dipole (O-H bond) and a component of polarizability anisotropy perpendicular to the O-H bond can be envisaged. In this unit cell, the total polarizability anisotropy is negative, $\Delta\Pi_{\text{trans}}^{(m)} < 0$, because $\Pi_{\text{trans}\parallel}^{(m)}$ corresponds to the restricted translational motion of a single water molecule. The expected TKE response of this supramolecule becomes negative.

The latter assignment of the H-bond bending dynamics as the origin of the TKE response of water is also endorsed by the Raman response of water. In the Raman spectrum of water,[13,51] the H-bond bending mode is fit by a Lorentzian with the resonance frequency of $\Omega_B/2\pi \approx 50$ cm$^{-1}$ and the damping rate of $\gamma_B \approx 115$ cm$^{-1}$. The relaxation of this damped motion may be resolved (e.g. in an optical Kerr effect experiment) as an exponential decay $e^{-\frac{t}{\tau}}$, with $\tau = (\frac{c\gamma_B}{2})^{-1} \approx 0.57$ps and $c$ being the speed of light. Although the interpretation of the OKE response of water has been controversial,[29,30,31,32,33] some authors also assigned an intermediate time constant of about 0.5ps to the H-bond bending of water[52,53]. Note also that, as the H-bond bending mode is not (or very weakly) IR active,[54,55] it is unlikely that the TKE response of water originates from a direct THz excitation of this mode.

We further studied the TKE response of water as function of temperature. We observed that the tail of the TKE signal relaxes faster by increasing the temperature (see Fig. S4). This finding is in line with the change of the bandwidth of the H-bond bending mode of water as function of temperature. In a Raman study, Mizoguchi et al. observed an increase of the bandwidth of this mode by increasing temperature.[56] Obviously, a larger bandwidth of a damped oscillator is resolved by a faster relaxation of its dynamic Kerr response.

Here we note that, it is quite rational that the O-O stretch vibration is also affected by the coupling to the rotational motions in H-bonded network of water, however, it may not be resolved in the TKE experiment because of our limited time resolution. For this mode, the induced birefringence is expected to relax with a time constant $(\frac{c\gamma_{O-O}}{2})^{-1} \approx 0.3\text{ps}^{36}$ (see Supplementary Note 2 and Fig. S5).

In summary, upon resonant excitation of the low-frequency rotational motion of water molecules, a Raman response is observed which is consistently ascribed to the restricted translational motion of water molecules. This response, which arises from the coupling of the intermolecular degrees of freedom of water, declares a pathway for the dissipation of external THz energy into the network of H-bonds. Our molecular dynamics simulations corroborate this conclusion and show the increase of the kinetic energy of the intermolecular translational motion after the initial coupling of the THz electric field to the rotational motions. The ultrafast flow of energy in the H-bonding network of water may be explained by the strong anharmonicity of the interaction energy of the intermolecular degrees of freedom. This approach opens a new avenue for using an intermolecular modes of water as local probes for the structural deformation of the H-bonding network of water in the vicinity of solutes, such as ions and biological macromolecules.

## Methods

**THz Kerr effect experiment.** For the TKE measurements, intense THz fields at ~1 THz are generated by a lithium Niobate ($LiNbO_3$) source with the tilted-pulse-front technique.[57] In the experiment, the linearly polarized THz pump pulse is focused onto the sample cell. The induced transient birefringence is measured by a temporally delayed and collinearly propagating probe pulse whose incident linear polarization is set to an angle of 45° relative to the THz electric field. Due to the pump-induced birefringence, the probe field components polarized parallel (∥) and perpendicularl (⊥) to the pump field acquire a phase difference $\Delta\phi$ when propagating through the sample, thereby resulting in elliptical polarization. The $\Delta\phi$ is detected with a combination of a quarter-wave plate and a Wollaston prism which splits the incoming beam in two perpendicularly polarized beams with power $P_\parallel$ and $P_\perp$. In the limit $|\Delta\phi| \ll 1$, the normalized difference $P_\parallel - P_\perp$ fulfills

$$\frac{P_\parallel - P_\perp}{P_\parallel + P_\perp} \approx \Delta\phi \quad (2)$$

and is measured by two photodiodes as a function of the temporal delay between THz pump and probe pulse[28].

For the temperature-dependent TKE measurements, the static cell is attached to a Peltier element and the temperature of the liquid is calibrated in advance. The stability and accuracy of the liquid's temperature is determined being ± 0.1 K.

**Water vapor.** The details of the water vapor experiment are given elsewhere[44]. Briefly, a gas jet of water molecules is excited by intense, linearly polarized THz pulses from the Lithium Niobate. The resulting transient optical birefringence is sampled similar to that of the liquid water. Water is kept inside a small vessel and heated up to 75 °C. The vapor is brought to the THz focus via a nozzle which has a hole with a diameter of 0.5 mm. The distance between the liquid surface and the THz focus is about 5 mm. Using a sucking tube (on top of the nozzle) we exit the vapor from the purging box. We estimate the thickness of the gas flow to be about 1 mm.

**Temperature rise.** To ensure that the accumulation of pump heat does not influence the results, we performed the TKE experiments also in a flow cell with the same SiN windows. We found no difference between static and flow cells in terms of both dynamics and amplitudes of the signals. Noteworthy that, the simple calculations based on $\Delta T = Q/mC$, with THz energy $Q \approx 6$ µJ, the mass of the excited volume of water $m \approx 3 \times 10^{-5}$ g and heat capacity of water $C = 4.18$ J/g estimate a negligible temperature rise $\Delta T \approx 50$ mK.

The AIMD results also confirms that the temperature rise and the change in the hydrogen bond density along the AIMD trajectory are negligible (see Fig. S6 and Fig. S7). The H-bond survival probability (Fig.

S8) also shows no effect of the THz excitation on the lifetime of a hydrogen bond. Therefore, the THz excitation in the experiment can be regarded as a small perturbation which minimally distorts the H-bonded structure of water. Interestingly, after the pulse, we find a slight increase in the probability of HB being broken due to translational diffusion of an initially H-bonded partner, and a slight decrease in the probability that a HB is broken because of rotational diffusion of a HB donor relative to the acceptor, with the effects cancelling each other so that the probability of survival of the H-bonds remains unaffected by the pulse.

*Ab initio* **Molecular dynamics simulations.** In this approach, the liquid is treated as a collection of nuclei and electrons to mimic the bulk conditions, under periodic boundary conditions. Molecular structure, interaction with the time-dependent electric field, and dielectric polarizability follow consistently from the electronic density of the quantum mechanical ground state as obtained within the Kohn-Sham formalism. Density functional theory-based *ab initio* MD simulations of a periodic cubic cell with 128 water molecules were performed at constant energy and ambient density (0.9966 g/cm$^3$) using the second generation Car–Parrinello method[58,59]. The energy and forces were computed using the mixed Gaussian-plane wave approach,[60] where the Kohn–Sham orbitals were represented by an accurate triple-$\zeta$ basis set with two sets of polarization functions (TZV2P),[61] and plane-waves with cutoff of 400 Ry were used to represent the charge density. The BLYP exchange-correlation functional plus a damped interatomic potential to account for van der Waals interactions (Grimme-D3[62]) were employed. Previous works have shown that this set-up provides a realistic description of many important structural, dynamical and spectroscopic characteristics of liquid water, including the partial pair correlation functions, self-diffusion and viscosity coefficients, HB lifetime, NMR shielding, as well as x-ray absorption and vibrational spectra[6,7,63]. The system was equilibrated for 30ps before 20 de-correlated snapshots were extracted from a further 40ps segment of the equilibrated trajectory. Each snapshot was then used to start an individual trajectory under the effect of the same THz pulse profile and amplitude as used in our experiment. No thermostat was employed and all the theoretical results and figures reported herein are averaged over the 20 trajectories. We used a Berry-phase approach to ensure a proper description of the field under periodic boundary conditions[64,65,66]. All computations were performed using the QUICKSTEP module of the CP2K suite of programs[67]. For defining a hydrogen bond we used a simple geometric criterion (3.5 Å and 30 degrees[68]).

**Classical Molecular dynamics simulations.** In the FFMD, the atomic pair interactions are parameterized using a combination of Lennard-Jones and Coulomb potentials. We use a rigid three-point water model (SPC/E) optimized to reproduce the density and structure of water.[69] We simulate a box with 5360 SPC/E water molecules in the NVE ensemble using the GROMACS version 2018.[69,70] We use 3D periodic boundary conditions, a Lennard-Jones potential which is shifted by a constant such that it is zero at the cut-off at 0.9 nm. 3D Particle Mesh Ewald summation are applied for the electrostatics beyond 0.9 nm. At time $t_0 = 5$ ps we apply a THz pulse according to

$$E(t) = \frac{1}{2v} E_0 \partial_t e^{-(t-t_0)^2/\sigma^2} \cos(v(t-t_0) + \varphi)$$

with $v = 0.5$/ps, $\sigma = 0.5$ ps = 0.5 ps, $\varphi = 0.628319$. We use $E_0 = 0.8$ MV/cm which is about three times higher than the experimental field amplitude, in order to have a good signal to noise ratio.

We calculate the total KE from the atomic velocities and the translational $KE_{trans}$ from the velocity of the molecular centers of mass. The $KE_{rot}$ is the difference between the two. During the simulation, the temperature increases on average less than 1 K. The curves shown in the paper are based on the average of 5000 trajectories of 50 ps, each containing a single pulse. We use a time step of 2 fs and a write-out frequency of 20 frames per ps.

We check for finite-size effects by simulating a water box of 128 molecules using the same procedure (see Fig. S9). As expected, the statistics are insufficient to draw any conclusions when only a small number of pulses is averaged, but for 3200 pulses and above, the curves reproduce the results for the larger box size. This shows that finite-size effects are unimportant.

**Kinetic energy decomposition.** In the AIMD simulations, at each MD snapshot the decomposition was done for every molecule in the instantaneous molecular internal coordinates defined by the Eckart conditions (the Eckart frame), thus minimizing the vibrational-rotational (Coriolis) cross term[71,72]. In the FFMD, the $KE_{rot}$ is calculated as the difference between the total KE and the KE of the molecular centers of mass.

**Anharmonic coupling**. Here, we suggest a model by which the flow of THz energy within the network of H-bonds can be rationalized. As shown schematically in **Fig. 6a**, two field interaction induces rotational anisotropy in the system. The energy of this interaction is rapidly (faster than our time resolution) transferred to the restricted translational motion of the neighboring water molecules and causes translational anisotropy in the system. Eventually, the induced anisotropy relaxes exponentially with $\tau \approx 0.5$ps. In this picture, the interaction energy of the rotational $Q_\theta$ and restricted translational $Q_B$ coordinates may be given by $V_1 = -\frac{g}{6} Q_B Q_\theta^2$, where $g$ is the anharmonic coupling constant. Under this coupling potential the equation of motion of $Q_B$ is expressed by[73,74,75]

$$\ddot{Q}_B + \gamma_B \dot{Q}_B + \Omega_B^2 Q_B = -\frac{\partial}{M \partial Q_B} V(t) = \frac{g}{6M} Q_\theta(t)^2 \qquad (3)$$

where $\Omega_B = 2\pi \nu_B$, $M$ is the mass of a water molecule and $Q_\theta(t)$ illustrates the dynamics of the rotational coordinate in which the THz field couples in. The dynamics of $Q_\theta(t)$ may be obtained from the diffusion equation with the driving force of $\mu_{eff} \times E_{THz}(t)$[22]. Therefore, the instantaneous energy transfer between the rotational motion and the H-bond bending mode of water may be explained by the large anharmonicity of their interaction energy $V_1$. Interestingly, by assuming that in the OKE process also the relaxation of the H-bond bending mode is observed, we can estimate the magnitude of the $g$ factor. We first start with solving Eq. (3). To this end, we need an estimate for $Q_\theta(t)^2$, the orientational motion which THz electric field couples in which can be obtained by solving the diffusion equation of the molecular orientational degrees of freedom, driven by the THz electric field torque. Since a small orientation is expected as the result of the THz torque, the molecular alignment $\langle \cos^2 \theta(t) \rangle$ is replaced by $Q_\theta(t)^2$, thereby[22,76]

$$Q_\theta(t)^2 \approx 3R_2 * [E \cdot \chi^{ori} * E]. \qquad (4)$$

Here, $E(t)$ is the amplitude of the linearly polarized pump field, $R$ illustrates the relaxation process, $*$ stands for the convolution operator i.e. $(\chi^{ori} * E)(t) = \int dt' \chi^{ori}(t-t')E(t')$ and $\chi^{ori}(t)$ is the contribution of the orientational molecular motion to the total dielectric susceptibility of water at the THz pump frequency. Since we observe an instantaneous energy transfer between the two coordinates (faster than the time resolution of the TKE experiment, ~350 fs), $R(t)$ is approximated by a delta function, and the driving force in Eq. (3) is simplified to $\frac{1}{2M} g \chi^{ori} E_{THz}^2 \delta(t)$.

The H-bond bending mode $Q_B$ can also be driven by an optical pulse through a Raman process, by which, the driven force is $\frac{N}{2M} \frac{\partial \Pi_{trans}}{\partial Q_B} E_{opt}^2 \delta(t)$,[77] where $N = 33 \times 10^{21}$ cm$^{-3}$ is the number density of water molecules.

By solving the equation of motion with the THz and the optical forces, we obtain respectively, $Q_B = \frac{1}{2M} g \chi^{ori} E_{THz}^2 \frac{\tau}{4\omega_1}$ and $Q_B = \frac{1}{M} N \pi^{1/2} \frac{\partial \Pi_{trans}}{\partial Q_B} E_{opt}^2 \frac{\tau}{4\omega_1}$, where $\omega_1 = \left( \Omega_B^2 - \frac{\gamma_B^2}{4} \right)^{1/2}$ and $\tau \approx 350$fs is the duration of the pulses[78]. We now compare the amplitude of the THz and optical induced birefringence at $t = 0$ and using the relation $\Delta n(t) = \frac{2\pi}{n} \frac{\partial \Pi_{trans}}{\partial Q_B} Q_B(t)$,[79] thus we obtain $\Delta n_{THz}/\Delta n_{OKE} = Q_{B_{THz}}/Q_{B_{opt}} = 3g\chi^{ori}/\frac{\partial \Pi_{trans}}{\partial Q_B} \approx 30$. In this equation, $\chi^{ori}$ is known form the dielectric spectrum of water and $\frac{\partial \Pi_{trans}}{\partial Q_B}$ is determined with the following procedure.

Upon propagation of the optical pump and probe pulses through the water cell with the thickness of $L = 100$ µm, the probe pulse (with wavelength λ) encounters a phase retardation $\Delta \phi$ proportional to the change in the refractive index of the sample: $\Delta n(t) \approx \frac{\lambda \Delta \phi(t)}{2\pi L}$. We find the maximum of $\Delta n$ from our experiment being $\Delta n_{\text{opt}} \approx 8 \times 10^{-9}$. Note that the $\Delta n_{\text{opt}}$ is obtained for the 0.5ps component without the contribution of the electronic response. $\Delta n_{\text{opt}}$ is also related to the nonlinear refractive index $n_2$ via $\Delta n = n_2 I = n_2 c \varepsilon_0 n_0 E^2$, $\varepsilon_0$ is the vacuum permittivity, $n_0$ is the linear refractive index.[35] With $I \approx 14 \text{ GW/cm}^2$ we obtain $n_2 \approx 30 \times 10^{-20} \text{ cm}^2/\text{W}$.

On the other hand, $n_2$ is related to the Raman scattering cross section $\sigma_{RS}$ via $\frac{n_0^2}{283} n_2 = \frac{2c^4 N \sigma_{RS}}{3\hbar \Omega_B^4 \gamma_B}$,[80,81] hence $\sigma_{RS} \approx 3 \times 10^{-33}$ cm². Finally using the relation $\sigma_{RS} = \omega_{\text{probe}}^4 \hbar \left(\frac{\partial \Pi_{\text{trans}}}{\partial Q_B}\right)^2 / 2M\Omega_B c^4$ [82] we obtain $\frac{\partial \Pi_{\text{trans}}}{\partial Q_B} \approx 3 \times 10^{-17} \text{cm}^2$ and accordingly,

$$g \approx \frac{10}{\chi^{\text{ori}}} N \frac{\partial \Pi_{\text{trans}}}{\partial Q_B} \approx 1 \times 10^{-10} \text{ J\AA}^{-1}\text{mol}^{-1}$$

**Acknowledgment**. We thank Takaaki Sato and Richard Buchner for the dielectric fit parameters of water. T.D.K. acknowledges funding from the European Research Council (ERC) under the European Union's Horizon 2020 research and innovation program (grant agreement No 716142). The generous allocation of supercomputer time by the Paderborn Center for Parallel Computing (PC²) is kindly acknowledged.